\documentstyle[preprint,version2,aps]{revtex}

\begin{document}
\draft
\begin{title}
\vskip2.5cm
Persistent Currents in Mesoscopic Hubbard Rings \\
with Spin-Orbit Interaction
\end{title}
\author{Satoshi Fujimoto and Norio Kawakami}
\begin{instit}
Yukawa Institute for Theoretical Physics,
Kyoto University, Kyoto 606, Japan
\end{instit}
\begin{abstract}
The effect of spin-orbit interaction on persistent
currents in mesoscopic Hubbard rings threaded
by an Aharonov-Bohm flux is investigated putting stress
on the orbital magnetism.  The non-perturbative treatment
of the spin-orbit interaction developed by Meir {\em et al.} is
combined with the Bethe ansatz solution to deal with
this problem exactly.  We find that the interplay of spin-orbit
interaction and electron-electron interaction plays a
crucial role, bringing about some new effects
on the orbital magnetism.
\end{abstract}
\vspace{1 cm}
\narrowtext
\section{INTRODUCTION}
Persistent equilibrium current occurring in isolated mesoscopic
normal metal rings penetrated by an Aharonov-Bohm (AB) flux is
one of the most interesting phenomena in mesoscopic systems
\cite{alt2,levy,chan,cheu,bou,ambe,alt,schmid,oppen}.
It is periodic in the flux with period of a flux quantum
(or half a flux quantum) and has either diamagnetic
or paramagnetic sign  according to different
experimental situations \cite{levy,chan}.
It has been widely accepted that spin-orbit (SO) interaction
may play a crucial role in the orbital magnetism, e.g.
the sign and the period of currents.  L\'evy {\em et al.}
have suggested that persistent currents observed by them
may have a diamagnetic sign due to the effect of strong
SO interaction (although they pointed out that their
determination contains some ambiguity)\cite{levy}. On the
other hand, Altshuler {\em et al.} have claimed from
thermodynamical arguments that the current averaged over
spatial disorder is always paramagnetic even in the strong SO
interaction limit \cite{alt}.  Extensive studies done subsequently
\cite{mathur,entin} have confirmed the result of
Altshuler {\em et al.}, and furthermore have revealed
some interesting aspects for SO effects
on the orbital magnetism.  Particularly in a non-perturbative
approach employed in \cite{entin} universal and
nonuniversal aspects of SO effects have been discussed,
such as the reduction factor of currents,
etc.  The above interesting studies
on SO effects, however, have been concerned with a free
electron model without electron-electron interaction.
It is hence desirable to examine how electron-electron interaction
is combined with SO interaction to affect
persistent currents in an AB geometry.

In this paper we wish to investigate the effect of
SO interaction on persistent currents in mesoscopic rings
of {\it mutually interacting} electrons for the canonical
ensemble.  For this purpose, we study the Hubbard ring with SO
interaction by combining  the non-perturbative treatment
of SO interaction of Meir {\em et al.}
\cite{meir} with the  Bethe ansatz technique\cite{woy,yu}.
We find that the interplay of SO interaction
and electron-electron interaction produce some
new effects on the orbital magnetism.
We further point out that simple reduction
factors of currents due to the SO effect
are modified in the presence
of electron-electron interaction.

The organization of the paper is as follows. In Sec. II, we
briefly depict how to diagonalize the  Hubbard ring
with SO interaction by the Bethe ansatz method, and
derive the excitation spectrum exactly. The key point is that the
non-perturbative treatment of SO interaction
for a non-interacting case\cite{meir} is
still applicable to the Hubbard model because of
local SU(2) symmetry of the Hubbard interaction.
We then study, in Sec. III, the SO effects
on persistent currents for case of the canonical
ensemble, putting stress on the sign and the period.
As is the case without SO interaction
\cite{yu,loss2}, the current shows a quite different
behavior according to the number of electrons
$N_c$ {\em modulo} 4.
We will see that upon averaging over  strong SO interaction,
some new effects on persistent currents are brought about
by the electron-electron interaction.
In Sec. IV, a possible extension to more
general models including long-range interaction is presented.
Summary and conclusion are given in Sec. V.

\section{HUBBARD RING WITH SPIN-ORBIT INTERACTION}

We consider the mesoscopic Hubbard ring with SO interaction.
The Hamiltonian reads
\begin{equation}
H=\sum_{i,\sigma,\sigma^{'}}
t_i(S_i)_{\sigma,\sigma^{'}}c^{\dagger}_{i,\sigma}
c_{i+1,\sigma^{'}}+h.c.+U\sum_{i}c^{\dagger}_{i\uparrow}
c_{i\uparrow}
c^{\dagger}_{i\downarrow}c_{i\downarrow}. \label{eqn:hamil}
\end{equation}
where $(S_i)_{\sigma,\sigma^{'}}$ is a matrix of SU(2),
and $t_i$ is chosen to be real for the sake of time reversal
symmetry.  The spin-dependent hopping matrix in (\ref{eqn:hamil})
reflects the effect of SO interaction.
The interplay between the Hubbard interaction and the
SO interaction makes it difficult to treat the model directly.
It is found, however, that one can still deal with the
Hamiltonian exactly by
combining the Bethe-ansatz technique with  the non-perturbative
approach developed for a non-interacting
case \cite{meir}. The point is that the hopping term
in (\ref{eqn:hamil}) with SO interaction is cast into the
diagonal form in spin space by
a unitary transformation
\begin{equation}
U^{(i)}=U^{(1)}S_1\cdot\cdot\cdot S_{i-1},  \label{eqn:unitary}
\end{equation}
with the matrix $U^{(1)}$ being chosen appropriately.
The transformation rotates the frame of spin space
by a different angle at each site so as to
produce the homogeneous hopping matrix.
An important point is that the onsite Coulomb term
in (\ref{eqn:hamil}) is invariant under such a local spin
rotation because of its SU(2) symmetry
\cite{fradkin}.  Hence one can formally gauge
away SO interaction, and find the
conventional Hubbard model in the new spin space,
\begin{equation}
H=\sum_{i,\tilde{\sigma}}t_i c^{\dagger}_{i,\tilde{\sigma}}
c_{i+1,\tilde{\sigma}}+h.c.
+U\sum_{i}c^{\dagger}_{i\tilde{\uparrow}}c_{i\tilde{\uparrow}}
c^{\dagger}_{i\tilde{\downarrow}}c_{i\tilde{\downarrow}},
  \label{eqn:hamil2}
\end{equation}
where $\tilde{\sigma}$ denotes a transformed spin variable,
and $\tilde{\uparrow}$ and $\tilde{\downarrow}$
label ``up'' and ``down''  spins respectively in the
new spin frame defined differently on each site.
The effect of SO interaction is now  incorporated
into spin-dependent twisted boundary conditions
for the eigenfunctions \cite{meir},
\begin{equation}
\psi_{\tilde{\sigma}}(N+1)=\exp[i(\Phi+\tilde{\sigma}\delta)]
\psi_{\tilde{\sigma}}(1),  \label{eqn:twistbc}
\end{equation}
where $\exp(i\tilde{\sigma}\delta)$ is the eigen value of
$S_1\cdot\cdot\cdot S_N$,
 $\tilde{\sigma}=\pm 1$ corresponds to ``up'' and ``down''
spins in the new spin space, and $\Phi$ is an AB flux
in unit of $\Phi_0/(2\pi)$ with the flux quantum $\Phi_0=h/e$.
The electron wave function acquires an additional phase
shift $ \pm \delta$ due to SO interaction after transversing the
ring. We will refer to $\delta$ as the SO phase shift hereinafter.
It is noteworthy that the above trick to simplify the
model is not specific to the Hubbard-type interaction,
but is  also applicable to any interaction,
including long-range type, which has local SU(2) symmetry.
Such extensions will be discussed later in this paper.

In the following, $t_n$ is assumed to be site-independent
so that the lattice is regular without
spatial disorder. Thus we put $t_i=-t$.
It is now straightforward to obtain the Bethe ansatz
solution to the Hamiltonian (\ref{eqn:hamil2})
with twisted boundary conditions (\ref{eqn:twistbc})
\cite{shastry}.
Following a standard method, two kinds of rapidities are
necessarily introduced to diagonalize the Hamiltonian.
For the ring system with $N$ sites, one thus gets the coupled
transcendental equations for charge ($p_j$) and
spin ($\lambda_\alpha$) rapidities \cite{lieb,shastry},
\begin{equation}
p_jN = 2\pi I_j + (\Phi +\delta)
+ \sum_{\beta=1}^{N_s} \eta (\sin p_j - \lambda_\beta),
 \label{eqn:bethe1}
\end{equation}
\begin{equation}
\sum_{j=1}^{N_c} \eta (\sin p_j - \lambda_\alpha)
                = 2\pi J_\alpha - 2 \delta
 - \sum_{\beta=1}^{N_s}\eta
((\lambda_\beta-\lambda_\alpha)/2),
  \label{eqn:bethe2}
\end{equation}
with $\eta(p)=-2{\rm tan}^{-1}(4tp/U)$,
where the number of total electrons (down-spin electrons)
is $N_c$ ($N_s$).  The quantum numbers $I_j$ and
$J_\alpha$ are integers (or half integers), which
specify charge and spin excitations.
The total energy is given in terms of the charge rapidity,
$E=-2t\sum_{j=1}^{N_c} \cos p_j$.
Note that the effect of SO interaction effectively
shifts  the quantum numbers $I_j$ and $J_\alpha$ whereas
that of Coulomb interaction appears via the
two-body phase shift function $\eta(k)$.

Let us now consider the effect of SO interaction
on the energy spectrum. Applying  a machinery
developed by Woynarovich  \cite{woy} to
Eqs.(\ref{eqn:bethe1}) and (\ref{eqn:bethe2}),
one can readily classify the excitation spectrum
including the finite-size corrections which are important
for the mesoscopic Hubbard ring.
For the fixed number of electrons, low-lying excitations
are specified by two kinds of quantum numbers $D_c$ and $D_s$,
which respectively carry the momentum $4k_FD_c$
(charge current) and $2k_FD_s$ (spin current), where $k_F$
is the Fermi momentum.  The excitation spectrum
including  SO interaction is written down as \cite{woy,frahm},
\begin{equation}
E(\Phi,\delta)-E_0=\frac{4\pi
v_c}{N}K_\rho \biggl(D_c+\frac{D_s}{2}+\frac{\Phi}{2\pi}\biggr)^2
+\frac{\pi v_s}{N}\biggl(D_s-\frac{\delta}
{\pi}\biggr)^2,
\label{eqn:energy}
\end{equation}
where $v_c$ and $v_s$ are the velocities of charge and spin
excitations respectively, and $K_\rho$
is the critical exponent for the  $4k_F$ oscillating
piece of charge correlation functions \cite{woy}.
For the Hubbard model $ 1/2 \leq K_\rho \leq 1$
\cite{frahm,shultz,kawa}.
These three fundamental quantities
characterize the Luttinger liquid properties of interacting
electrons completely \cite{hal},
which can be straightforwardly evaluated using the Bethe-ansatz
integral equation resulting from (5) and
(6)\cite{woy,frahm,shultz,kawa}.
For a given number of electrons it is necessary to
find the lowest energy state  correctly
in order to derive the expression for currents.
It is crucial for this purpose to notice that
the quantum numbers $D_c$ and $D_s$ respect the
following selection rule reflecting the Fermi statistics
\cite{woy},
\begin{equation}
D_c=\frac{N_c+N_s+1}{2} \qquad \mbox{(mod 1)},
  \label{eqn:rule1}
\end{equation}
\begin{equation}
D_s=\frac{N_c}{2} \qquad \mbox{(mod 1)}.
  \label{eqn:rule2}
\end{equation}
We note that in the absence of SO interaction, the above
spectrum has been analyzed  by Yu and Fowler
to study persistent currents \cite{yu}.

\section{EFFECTS OF SPIN-ORBIT INTERACTION ON PERSISTENT CURRENTS}

We now study the effects of SO interaction on
persistent equilibrium currents in the Hubbard ring.
Since we wish to discuss the
magnetism for the canonical ensemble with the fixed number of
electrons, let us briefly summarize some results known for
the canonical ensemble.
The property of the orbital magnetism is sensitive to the
number of fermions carrying the currents\cite{byers,loss}.
For example,
a free electron model without SO interaction the orbital
magnetism
depends on the total number of electrons $N_c$
{\em modulo} 4;
{\em i.e.}, the ground state is diamagnetic for $N_c=4n+2$,
paramagnetic for $N_c=4n$ with period of a
flux quantum, and paramagnetic with period
of half a flux quantum for $N_c=4n+1, 4n+3$ \cite {loss2}.
The results are slightly modified in the presence
of electron-electron interaction as shown
for the 1D Hubbard model \cite{yu,fye}. For instance
the paramagnetic state for $N_c=4n$ is altered
to a diamagnetic one by electron-electron
interaction except near half filling (one electron
per lattice site).
In the presence of SO interaction, further modifications
are expected to occur in the orbital magnetism
due to the interplay of SO interaction and
electron-electron interaction. In particular
we will see below that for a certain  parameter regime
of the interaction strength,
the paramagnetic state is stabilized for $N_c=4n+2$
in contrast to the case  without SO interaction
for which the ground state is always diamagnetic.

In order to
clearly see what is going on, we first study
the case of $4n+2$ in detail, and mention the other
cases later in this section. It may be plausible
to introduce here an important key quantity
$v_cK_{\rho}/v_s$, which will be
helpful for following arguments.  For noninteracting
electrons,  $v_cK_{\rho}/v_s=1$ for any electron concentrations
because  $v_c=v_s$ and $K_\rho=1$, whereas in  correlated cases
of $U \neq 0$ the value of  $v_cK_{\rho}/v_s $
ranges from $0$ to $\infty$ depending on the interaction
strength as well as electron concentrations.
We show $v_cK_{\rho}/v_s $  as a function of
electron concentrations in Fig. 1.  Roughly speaking,
one can see from this quantity whether
the SO effect may be enhanced
or suppressed by the electron-electron interaction.  For example,
the SO effect is suppressed for  $v_cK_{\rho}/v_s >1$,
whereas it is enhanced for $v_cK_{\rho}/v_s <1$.

\subsection{Persistent currents for $N_c=4n+2$ $(N_s=2n+1)$}
It is seen from Eqs.(\ref{eqn:rule1}) and (\ref{eqn:rule2})
that the selection rule for the quantum numbers in this case
is $D_c=0$ (mod 1) and $D_s=0$ (mod 1), which
implies that the ground state in the absence of
SO interaction is diamagnetic around
$\Phi=0$\cite{shultz,fye,kawa2}.
Turning on an AB flux, there occur two different
situations according to the magnitude of
$v_cK_{\rho}/v_s$.  It is straightforward to
derive the persistent current $I=-\partial
E/\partial \Phi$ from Eqs. (\ref{eqn:energy}).
For $v_cK_{\rho}/v_s \geq 1$, the current takes the form\cite{yu}
\begin{equation}
I=\left\{
   \begin{array}{ll}
 \displaystyle{-\frac{v_cK_{\rho}}{\pi N}(\Phi+\pi)}&
     \qquad -\pi\leq\Phi\leq -\Phi_c \\
    \noalign{\vskip0.2cm}
 \displaystyle{-\frac{v_cK_{\rho}}{\pi N}\Phi}&
     \qquad -\Phi_c < \Phi < \Phi_c \\
    \noalign{\vskip0.2cm}
 \displaystyle{-\frac{v_cK_{\rho}}{\pi N}(\Phi-\pi)}&
     \qquad \Phi_c\leq\Phi\leq\pi
   \end{array} \right.   \label{eqn:cur1}
\end{equation}
where
\begin{equation}
\Phi_c=\frac{\pi}{2}+\frac{\pi v_s}{2K_{\rho} v_c}
-\frac{v_s\delta} {K_{\rho}v_c}.
\label{eqn:phi}
\end{equation}
One can see that the SO phase shift $\delta$ simply alters the
critical value $\Phi_c$ of the AB flux, and
the diamagnetic nature around $\Phi=0$ is not modified by SO
interaction. From Eq.(\ref{eqn:phi}) it is explicitly seen
that the effect of SO interaction is suppressed
as the value of $v_cK_{\rho}/v_s$ increases.
Such cases with  $v_cK_{\rho}/v_s \geq 1$
realize at lower electron concentrations for $U \neq 0$,
as seen from Fig. 1. It is instructive to point out here
that for $\delta=\pi/2$
the effects of electron-electron interaction
disapear in Eq.(\ref{eqn:phi}), and hence
the period is halved as has been known for a free
electron model\cite{meir}.  An alternative expression of
Eq.(\ref{eqn:cur1}) in Fourier series expansion is found
to be more convenient for following discussions,
\begin{equation}
I(\Phi)=\sum_n\frac{v_cK_{\rho}}{\pi N}\frac{(-1)^n}{n}
\cos\biggl[n\biggl(\frac{\pi}{2}-\frac{\pi
v_s}{2K_{\rho}v_c}+\frac{v_s\delta}{K_{\rho}v_c}\biggr)\biggr]
\sin(n\Phi).  \label{eqn:curf1}
\end{equation}

In contrast to the above case, the expression for
currents in the case of $v_cK_{\rho}/v_s< 1$
 depends on the value of the SO phase shift $\delta$.
For $0\leq\delta<\pi K_{\rho} v_c/2v_s+\pi/2$ the current
is given by the same expression as Eq.(\ref{eqn:cur1}),
whereas  for $\pi K_{\rho} v_c/2v_s+\pi/2\leq\delta\leq
\pi$ it is changed to
\begin{equation}
I =\left\{
   \begin{array}{ll}
 \displaystyle{-\frac{v_cK_{\rho} }{\pi N}(\Phi+\pi)}&
     \qquad -\pi\leq\Phi\leq 0 \\
    \noalign{\vskip0.2cm}
  \displaystyle{-\frac{v_cK_{\rho}}{\pi N}(\Phi-\pi)}&
     \qquad 0<\Phi\leq\pi .
   \end{array} \right.    \label{eqn:cur2}
\end{equation}
This is cast into an alternative formula in the
Fourier series expansion as
\begin{equation}
I(\Phi)=\sum_n\frac{v_c K_{\rho}}{\pi Nn}\sin(n\Phi).
\label{eqn:curf2}
\end{equation}
One can see from  Eq.(\ref{eqn:cur2}) that the current has
a paramagnetic sign around $\Phi=0$.
This change in sign is due to the
interplay of the SO interaction and
electron-electron interaction, which still plays a crucial role
upon averaging over the strong SO interaction (see below).

\subsection{Strong SO interaction limit}
We have seen  that the SO phase shift $\delta$
crucially modifies the characteristic behavior
of the orbital magnetism in correlated electrons.
The quantity $\delta$ is  to be determined in
the  range $[0, \pi]$ for a given scattering process
by SO interaction. For example
when the scattering length of SO interaction would be
comparable to the length of the ring, the value of  $\delta$
could be around $\pi/2$,
which results in the period of currents halved by the
effect of the SO interaction as seen from (\ref{eqn:curf1}).
In more general cases when the length of the ring may be
much larger than the scattering length,  one has to take into
account all possible rotations in spin space due to SO scattering.
Such a limit is referred to as the strong SO
interaction limit\cite{entin,meir}.
Since the SO phase shift $\delta$ is directly
related to the rotation angle in spin space,
one should average  currents over $\delta$
from $0$ to $\pi$ with the weight $\sin^2{\delta}
$\cite{entin,bergman}.  The Fourier transformed
formulae of currents are more convenient to
carry out the average. Using Eqs.(\ref{eqn:curf1})
and (\ref{eqn:curf2}), we thus  derive the expression
for persistent currents in the strong SO interaction limit.
In the case of  $v_cK_{\rho}/v_s \geq 1$
(low electron densities), we get
\begin{equation}
I(\Phi)=\sum_m\frac{v_cK_{\rho}}{2\pi^2 N}
\biggl(\frac{v_cK_{\rho}}{v_sm}+
\frac{v_sv_cK_{\rho}m}{4v_c^2K_{\rho}^2-4v_s^2m^2}\biggr)
\frac{(-1)^m}{m}
\sin\biggl(m\frac{v_s\pi}{v_cK_{\rho}}\biggr)\sin(2m\Phi).
\label{eqn:curfst1}
\end{equation}
It is seen that all the odd harmonics are dropped, and hence
the period is reduced to half a flux quantum. In contrast
to a free electron case in which only harmonics
$n=0, 2$ remain and the harmonics $n=2$ gives
a diamagnetic sign change with a reduction factor $1/2$
\cite{entin},
all higher even harmonics survive in the correlated case.
However it is found that  the ground state is still
{\it diamagnetic}
around $\Phi=0$. To see this more explicitly we
evaluate the sum of the coefficient by  Fourier transformation:
\begin{equation}
\sum_m \biggl(\frac{v_cK_{\rho}}{mv_s}+
\frac{v_sv_sK_{\rho}m}{4v_c^2K_{\rho}^2-4v_s^2m^2}\biggr)
(-1)^m\sin\biggl(\frac{v_sm\pi}{v_cK_{\rho}}\biggr)=-
\frac{1}{2},
\end{equation}
which results in  the diamagnetic
current $I(\Phi)=-(v_cK_{\rho}/\pi N)\Phi$. Note that
this is exactly same as the current without SO interaction.
Hence we can conclude that in case of $v_cK_{\rho}/v_s\geq 1$,
the behavior of currents around $\Phi=0$
is not modified by the SO effects
even in the strong interaction limit.
This is because the SO interaction can affect
the occupation of energy levels only
when there exists a finite AB flux in the system.

The situations are somewhat different for $v_cK_{\rho}/v_s< 1$
(close to half filling). There are both diamagnetic
and paramagnetic contributions to currents depending
on $\delta$. The current average over the strong SO
interaction is thus  given by
\begin{eqnarray}
I(\Phi)&=&\sum_n\frac{v_cK_{\rho}}{2\pi^2 N
n}\biggl[\frac{v_s^2n^2}{4v_c^2K_{\rho}^2-v_s^2n^2}
\sin\biggl(\frac{\pi v_cK_{\rho}}{v_s}\biggr)
+\frac{8v_c^3K_{\rho}^3}{v_sn(4v_c^2K_{\rho}^2-v_s^2n^2)}
\sin\biggl(\frac{n\pi}{2}+\frac{n\pi v_s}{2v_cK_{\rho}}\biggr)
 \nonumber\\
& &+\pi-\frac{\pi v_c K_{\rho}}{v_s}\biggr]
\times\sin(n\Phi). \label{eqn:curfst2}
\end{eqnarray}
The period is not halved upon averaging over $\delta$
in this case. For small positive $\Phi$ with
$1/3<v_cK_{\rho}/v_s<1 $,
one can perform the inverse Fourier transformation
to get for $\Phi\rightarrow 0$,
\begin{equation}
I(\Phi)=-\frac{2v_cK_{\rho}}{\pi N}\biggl\{\biggl
[-\frac{v_cK_{\rho}}{v_s}
\cos\biggl(\frac{\pi
v_cK_{\rho}}{v_s}\biggr)+\frac{v_s-v_cK_{\rho}}{v_s}\biggr]\Phi
-\frac{1}{2}\biggl[\pi-\frac{\pi v_cK_{\rho}}{v_s}-
\sin\biggl(\frac{\pi v_cK_{\rho}}{v_s}\biggr)\biggr]\biggr\},
\end{equation}
which has a  paramagnetic sign. Since paramagnetic
contributions to the current averaged
over $\delta$ increase compared with
diamagnetic ones when $v_cK_{\rho}/v_s$
is decreased, we can say that the current
for $0\leq v_cK_{\rho}/v_s<1$ always shows a
{\it paramagnetic} sign  around $\Phi=0$.
We would like to stress here that
this paramagnetic state realizes as a consequence
of the interplay of SO interaction and
electron-electron interaction, and should not appear
if either of the two is absent.

We have computed persistent currents using the formulae
(\ref{eqn:curfst1}) and (\ref{eqn:curfst2})
together with the Bethe equations for $v_c$,$v_s$, and
$K_{\rho}$\cite{woy}.
In Figs. 2 and 3, persistent currents in the strong
SO limit are shown as a function of the AB flux $\Phi$
for two different cases.

\subsection{Reduction factors in currents}
Here we make a brief comment on reduction factors in persistent
currents by the SO effects. According to Meir {\em et al.}
\cite{meir} and Entin-Wohlman {\em et al.}\cite{entin},
reduction factors due to the SO interaction
are written in a simple and universal form for
a noninteracting model with even number of electrons.
An essence of their idea is that the persistent current
$I(\Phi)$ in the presence of SO interaction
can be expressed in the form
\begin{equation}
I(\Phi)=I_0(\Phi+\delta)+I_0(\Phi-\delta)
\end{equation}
where $I_0$ is the current per each spin component
 without SO interaction. This leads to universal reduction
factors in the form of Fourier expansion \cite{entin,meir},
\begin{equation}
I(\Phi)=\sum_{n} \cos(n \delta) a_n \sin(n\Phi)
\end{equation}
with $a_n$ being the Fourier coefficients for the case
without SO interaction.  One can see that the reduction factor
$\cos(n \delta)$ depends only on the SO
phase shift.  This expression
leads to a rather simple result that upon
averaging in the strong SO limit, the current
is reduced by a factor of 1/2, and the fluctuations
by a factor 1/4 \cite{entin,meir}.
For interacting electrons, however, it is
seen that the SO phase shift should not appear
in a simple form of $\Phi \pm \delta$. It is instead
combined with a factor of $v_c K_\rho/v_s$
reflecting  the electron correlation effects.
Thus in the presence of electron-electron interaction,
the reduction factors follow from Eqs.(\ref{eqn:curf1}),
\begin{equation}
\cos\biggl[n \biggl(\frac{\pi}{2}-\frac{\pi v_s}{2v_cK_{\rho}}
+\frac{v_s\delta}{v_cK_{\rho}}\biggr)\biggr]/
\cos\biggl[n \biggl(\frac{\pi}{2}-\frac{\pi v_s}{2v_cK_{\rho}}
\biggr)\biggr], \label{eqn:reduction}
\end{equation}
for $v_cK_{\rho}/v_s\geq 1$. For $v_cK_{\rho}/v_s< 1$
we can not define the reduction
factor in such a simple form, as seen from
Eqs.(\ref{eqn:curfst2}).
So, the above remarkable properties obtained for the SO effects
based on a free electron model are changed when the
electron-electron interaction is introduced.
To avoid confusions we would like to mention that
the present nonuniversal results  may not
be contradicted with the universal reduction factor
of currents expected in disordered
system\cite{mathur,entin,bergman,hikami,alt3,lee}. In such
 cases the average over disorder may play an essential role,
which has not been taken into account in the
present calculation.

\subsection {Electron-number dependence }
We have been concerned so far with the case of
$N_c=4n+2$.  As mentioned before the results are
sensitively dependent on the number of electrons.
Here we summarize the results for other cases.
The calculation  can be performed in  parallel
to the above example of $N_c=4n+2$.

($a$) $N_c=4n$ ($N_s=2n$).
In this case the selection rule for the quantum
numbers reads, $D_c=1/2$ (mod 1) and $D_s=0$ (mod 1).
For $v_cK_{\rho}/v_s\geq 1$ the persistent current in the
Fourier transformed form is then given by
\begin{equation}
I(\Phi)=\sum_n\frac{v_cK_{\rho}}{\pi Nn}\cos\biggl[
n\biggl(\frac{\pi}{2}-\frac{v_s\pi}{2v_cK_{\rho}}+
\frac{v_s\delta}{v_cK_{\rho}}\biggr)\biggr]\sin(n\Phi),
 \label{eqn:curf3}
\end{equation}
which shows a diamagnetic sign around $\Phi=0$.
In the case of $v_cK_{\rho}/v_s< 1$, where only
the paramagnetic state realizes without SO interaction,
the expression is the same as Eq.(\ref{eqn:curf3})
for $\pi/2-v_cK_{\rho}\pi/2v_s<\delta\leq\pi$, while
 Eq.(\ref{eqn:curf2})  for $0\leq \delta\leq
\pi/2-v_cK_{\rho}\pi/2v_s$. Thus the paramagnetic
state near the metal-insulator transition
(half-filling) is changed to a diamagnetic one
by SO interaction for $\pi/2-v_cK_{\rho}\pi/2v_s<\delta\leq\pi$.
It is remarkable, however, that in the strong SO
interaction limit we obtain exactly the same results
in currents as for the case of $N_c=4n+2$
upon averaging over all possible phases $\delta$
(see Figs. 2 and 3).

We can hence summarize the results for
the  {\it even} number of electrons
as follows;  the ground state is always
diamagnetic for $v_cK_{\rho}/v_s\geq 1$
(low electron densities) in the strong SO limit,
whereas for $v_cK_{\rho}/v_s< 1$ (near the
metal-insulator transition point)
the paramagnetic state is stabilized not
only for $N_c=4n$ but also for $N_c=4n+2$ as a
result of the SO effects.

({\em b}) {\em Odd case} ($N_c=4n+1, 4n+3$).
We first consider the case for $N_c=4n+3$.
Since Eq.(\ref{eqn:energy}) holds
only for the case with  $N_{\uparrow}\geq N_{\downarrow}$,
the roles of ``up'' spin and ``down'' spin for $\Phi>0$
are interchanged for $\Phi<0$ in Eq.(\ref{eqn:energy}).
Thus the result for $\Phi<0$ in the case of
odd number of electrons can be deduced  from
that for $\Phi>0$ by changing  $\delta\rightarrow -\delta$.
Noting the selection rule $D_c=1/2$ (mod 1) and $D_s=1/2$ (mod 1)
for $\Phi>0$, the current in the ground state is now given by
\begin{equation}
I=\left\{
   \begin{array}{ll}
  \displaystyle{-\frac{v_cK_{\rho}}{\pi N}\biggl
(\Phi+\frac{\pi}{2}\biggr)}&
    \qquad -\pi\leq \Phi <0 \\
          \noalign{\vskip0.2cm}
  \displaystyle{-\frac{v_cK_{\rho}}{\pi N}\biggl
(\Phi-\frac{\pi}{2}\biggr)}&
    \qquad 0\leq \Phi\leq \pi \\
   \end{array} \right. \label{eqn:curodd1}
\end{equation}
which has a period of half a flux quantum, and exhibits
a paramagnetic sign around $\Phi=0$.
It should be noted that the above expression is independent
 of $\delta$, and hence there are not any modifications
due to SO interaction. The effect of SO interaction
indeed appears in  the next order corrections
in $1/N$, which are not taken into account in our formulation.
For a non-interacting case the next-order corrections
have been evaluated by Entin-Wohlman {\em et al.},
resulting in a small shift of energy minima
from $\Phi=\pm\pi/2$ to $\Phi=\pm[(1+1/N_c)\pi/2-\delta/N_c]$
\cite{entin,comment1}, which may be actually neglected in
mesoscopic rings.  We note that for another odd case of
$N_c=4n+1$,  we get the exactly  same results
of currents as for $N_c=4n+3$.

In summary we can say that characteristic
properties of persistent currents  in the strong SO limit
are classified  by the  parity  of the
electron number even for correlated electron systems.

\section{EXTENSION TO MORE GENERAL MODELS}

In the Hubbard model the electron-electron
interaction  is assumed to be short-ranged (on-site),
which in turn enables us to treat the model exactly.
In more general cases, however, effective interaction
would be of long-range type, since the screening
effect of the Coulomb interaction may become
less effective for mesoscopic metallic rings.
In such cases it is quite difficult to get the energy spectrum
exactly.  So it is desirable to find a possible way
to extend our analysis to more general cases including
long-range interaction. We wish to briefly depict
a simple idea how to treat such cases.

We recall here a trick used to simplify the Hubbard model
with SO interaction, i.e. a unitary
transformation which incorporates the
SO interaction into the boundary effects (\ref{eqn:unitary}).
Note that this technique is still applicable to
more general long-range interactions so long as they
retain local SU(2) symmetry. Such a local SU(2) symmetry
for interaction may be expected to hold in ordinary cases,
such as partially screened Coulomb interaction, etc.
The remaining problem is then how to obtain the
expression for the low-energy  spectrum like
Eq.(\ref{eqn:energy}) including the SO
effect and the AB flux.  To this end the bosonization
technique may be useful\cite{emery,solyom,shultz,hal}
because it can formally  describe
low-energy states even for nonintegrable systems.
The bosonization scheme has been
previously  used by Loss to discuss
persistent currents for a spinless fermion
system  \cite{loss}.

Following a standard technique in bosonization
\cite{emery,solyom}, we now  discuss how
the SO effect on currents can be treated in  nonintegrable
systems.   We do not have to specify an explicit form
of interaction here, but  only assume the interaction
$V(r)$ to be invariant under local SU(2) transformations.
In general, low-energy gapless states of 1D metallic
electron systems compose of two  independent
Luttinger liquids \cite{hal} corresponding to
charge and spin degrees of freedom \cite{frahm,shultz,kawa}.
Hence the system is described by the sum of
two Gaussian models with conformal charge $c=1$.

Let us now introduce the boson fields for spin and
charge degrees of freedom.
It is found that the spin-dependent twisted
boundary conditions (\ref{eqn:unitary}) due to an
AB flux and SO interaction
are incorporated into boson phase fields as \cite{hal,loss}
\begin{equation}
\phi_{\rho}=\sum_{k\neq 0}\biggl|\frac{\pi}{2Lk}
\biggr|^{\frac{1}{2}}
e^{ikx}[a_{k, \rho}^{\dagger}+a_{-k, \rho}]
+\phi_{0,\rho}+M_{\rho}\frac{\pi x}{L},
\end{equation}
\begin{equation}
\phi_{\sigma}=\sum_{k\neq 0}\biggl|\frac{\pi}{2Lk}
\biggr|^{\frac{1}{2}}
e^{ikx}[a_{k, \sigma}^{\dagger}+a_{-k, \sigma}]
+\phi_{0,\sigma}+M_{\sigma}\frac{\pi x}{L},
\end{equation}
\begin{equation}
\theta_{\rho}=\sum_{k\neq 0}\biggl|\frac{\pi}{2Lk}\biggr|
^{\frac{1}{2}}{\rm sgn}(k)
e^{ikx}[a_{k, \rho}^{\dagger}-a_{-k, \rho}]
+\theta_{0,\rho} +\biggl(J_{\rho}+\sqrt{2}\frac{\Phi}
{\pi}\biggr)\frac{\pi}{L}\biggl(x+\frac{L}{2}\biggr),
\end{equation}
\begin{equation}
\theta_{\sigma}=\sum_{k\neq 0}\biggl|\frac{\pi}{2Lk}\biggr|
^{\frac{1}{2}}{\rm sgn}(k)
e^{ikx}[a_{k, \sigma}^{\dagger}-a_{-k, \sigma}]
+\theta_{0,\sigma} +\biggl(J_{\sigma}
+\sqrt{2}\frac{\delta}{\pi}\biggr)\frac{\pi}{L}\biggl(x+\frac{L}{2}\biggr),
\end{equation}
where $a_{k,\rho}(a_{k,\rho}^{\dagger})$ and
$a_{k,\sigma}(a_{k,\sigma}^{\dagger})$
are boson annihilation (creation) operators for charge and spin
densities, respectively,
 and $M_{\rho} (M_{\sigma})$ and
$J_{\rho}$ ($J_{\sigma}$) are the charge (spin) number and
the charge (spin) current, respectively, which are defined by
$M_{\rho}=(M_{\uparrow}+M_{\downarrow})/\sqrt{2}$,
$M_{\sigma}=(M_{\uparrow}-M_{\downarrow})/\sqrt{2}$,
$J_{\rho}=(J_{\uparrow}+J_{\downarrow})/\sqrt{2}$,
and $J_{\sigma}=(J_{\uparrow}-J_{\downarrow})/\sqrt{2}$.
We note here that $M_{\uparrow(\downarrow)}$
and $J_{\uparrow(\downarrow)}$ describe
topological excitations introduced by Haldane\cite{hal}.
They satisfy the selection rules, $(-1)^{N_{0 s}+1}
=(-1)^{M_s+J_s}$, where $s=\uparrow$, $\downarrow$
and $N_{0 \uparrow(\downarrow)}$ is the total number
of up (down) spins.
$\phi_{0, \nu}$ and $\theta_{0, \nu}$ are conjugate variables of
$J_{\nu}$ and $M_{\nu}$, respectively.

Using the above phase fields we can write down
the low-energy effective  Hamiltonian,
from which the persistent current directly follows,
as long as  the system belongs to the universality class of
Luttinger liquids.
The energy spectrum of the effective Hamiltonian with
finite-size correction terms reads \cite{hal}
\begin{equation}
E(\Phi)-E_0=\frac{\pi}{2L}\biggl[v_{\rho M}M_{\rho}^2+v_{\rho J}
\biggl(J_{\rho}+\sqrt{2}\frac{\Phi}{\pi}\biggr)^2
+v_{\sigma M}M_{\sigma}^2+v_{\sigma J}
\biggl(J_{\sigma}+\sqrt{2}\frac{\delta}{\pi}\biggr)^2\biggr],
\label{eqn:lute}
\end{equation}
where $v_{\nu M}$ and $v_{\nu J}$ ($\nu=\rho$, $\sigma$)
are Luttinger liquid parameters describing the
velocity of excitations. It is to be noted that
the effect of electron-electron
interaction $V(r)$ is only to renormalize these parameters.
The ground state is given by the condition, $M_{\rho}=
M_{\sigma}=0$, {\em i.e.} $M_{\uparrow}=M_{\downarrow}=0$.
Thus using the selection rules mentioned above we obtain the
topological constraints for $J_{\uparrow}$ and $J_{\downarrow}$:
{\em (1)} $J_{\uparrow}$ even, $J_{\downarrow}$ even
for $N_c=4n+2$, {\em (2)} $J_{\uparrow}$ odd, $J_{\downarrow}$
odd for $N_c=4n$, {\em (3)} $J_{\uparrow}$ odd, $J_{\downarrow}$
even (and vice versa) for $N_c=4n+3$,
{\em (4)} $J_{\uparrow}$ even, $J_{\downarrow}$ odd (and vice
versa)
for $4n+1$. Consequently, if we put $v_{\rho J}=K_{\rho}v_c$
and $v_{\sigma J}=v_s$, we find that the energy
(\ref{eqn:lute}) with these topological constraints
is equivalent to Eq.(\ref{eqn:energy}),
and $D_c$ and $D_s$ correspond to $J_{\uparrow}/2$
and $(J_{\downarrow}-J_{\uparrow})/2$, respectively.

Using Eq.(\ref{eqn:lute}) and the above topological constraints,
we can obtain the persistent current, and discuss its sign and
period quite similarly to the case for the Hubbard ring.
Therefore all the expressions derived in the previous section
can be directly applied to the present case by regarding
$v_cK_{\rho}/v_s$ as a free parameter to be determined.
In order to determine the parameter,
we generally need input data from another microscopic
calculations, {\em e.g.} numerical diagonalization.
We note that several elegant techniques to get the
Luttinger liquid  parameters numerically for nonintegrable
systems have been already developed \cite{hal,shultz}, which
will be helpful for us to evaluate the persistent currents
explicitly.

\section{SUMMARY}

In this paper we have discussed the effects of SO interaction
on persistent currents in the mesoscopic Hubbard ring.
We have investigated the problem exactly
combining the Bethe-ansatz solution with a
unitary transformation which incorporates
the SO effects into spin-dependent twisted
boundary conditions. It has been shown that
characteristic properties of the orbital magnetism in the
Hubbard ring are classified according to
 the value of $v_cK_{\rho}/v_s$ and the number
of electrons {\em modulo} 4.  In particular, we
have demonstrated that $v_cK_{\rho}/v_s$ is
an important key quantity  to see whether the
SO effects are enhanced or suppressed by the
electron-electron interaction.

In the strong SO interaction limit it has been found
that the formula obtained for currents is classified by
the parity of the electron number.  For the even
number of electrons,  the ground state is
diamagnetic with period of half a flux quantum
for $v_cK_{\rho}/v_s\geq 1$ (low electron densities),
and paramagnetic with period of a flux quantum
for $v_cK_{\rho}/v_s<1 $ (close to half filling).
In particular the paramagnetic state for $v_cK_{\rho}/v_s<1$
is realized by a combined effect arising from
the interplay of SO interaction and electron-electron
interaction.  In the Hubbard model, the condition
$v_cK_{\rho}/v_s<1$ is  satisfied near half filling
which implies that the system would be close to the
Mott insulator. Therefore such a novel phenomenon for
$v_cK_{\rho}/v_s<1$ is expected in general to occur
for interacting electrons in a metallic phase
close to the Mott insulator. In contrast to the even case,
the ground state for the odd number of electrons is found to be
always paramagnetic with period of half a flux quantum,
which is not affected by SO interaction as long as
the corrections up to the order of $O(1/N)$ are concerned.

In conclusion the effect of SO interaction
together with  electron-electron interaction
gives rise to a novel and qualitative change
in the orbital magnetism for 1D interacting electron systems.
There remain  several important problems to be investigated.
For example we have not considered the effect of
disorder in this paper, which would
induce interesting phenomena together with
the SO effects as well as with the correlation effects.
Furthermore an extension of the theory
to multichannel cases is  desirable to
confront the results with various experiments.
These problems are now under consideration.

\nonum
\section{ACKNOWLEDGMENTS}

This work is partly supported by the Grant-in-Aid from the
Ministry of Education, Science and Culture.

\newpage


\newpage
\figure{  \label{Fig.1}}
Plots of  $v_cK_\rho/v_s$
as a function of electron densities $n$
for the Hubbard model.  The half filling corresponds
to $n=1$.
\figure{  \label{Fig.2}}
Persistent currents plotted against $\Phi/2\pi$
for $U/t=4$ and $n=0.65$ ($v_cK_{\rho}/v_s=1.32$)
in the case of even number of electrons.
The current normalized by $t/N$
is shown.
\figure{  \label{Fig.3}}
Persistent currents plotted against $\Phi/2\pi$
for $U/t=4$ and $n=0.95$
($v_cK_{\rho}/v_s=0.58$) in the case of
even number of electrons.
The current normalized by $t/N$
is shown.
\end{document}